\documentclass[12pt,reqno]{article}
\usepackage[a4paper]{geometry}
\usepackage{setspace}
\usepackage[utf8]{inputenc}
\usepackage[authoryear]{natbib}
\usepackage{url}
\usepackage{graphicx}
\usepackage{hyperref}
\hypersetup{
    colorlinks=true,
    linkcolor=red,
    citecolor=blue,
    filecolor=magenta,      
    urlcolor=cyan
    }

\onehalfspace

\begin{document}
    
\title{Racial and income-based affirmative action in higher education admissions: lessons from the Brazilian experience}

\author{Rodrigo Zeidan\thanks{NYU Shanghai, China and Fundação Dom Cabral, Brazil. Email: rz25@nyu.edu} \and Silvio Luiz de Almeida\thanks{Department of Philosophy and General Theory of Law, Fundação Getúlio Vargas, Brazil. Email: silvio.almeida@fgv.br} \and Inácio Bó\thanks{Southwestern University of Finance and Economics, China. Email: mail@inaciobo.com} \and
Neil Lewis, Jr.\thanks{Department of Communication, Cornell University, United States. Email: nlewisjr@cornell.edu}
}

\date{April 2023}

\maketitle

\begin{abstract}
This survey article provides insights regarding the future of affirmative action by analyzing the implementation methods and the empirical evidence on the use of placement quotas in the Brazilian higher education system. All federal universities have required income and racial-based quotas in Brazil since 2012. Affirmative action in federal universities is uniformly applied across the country, which makes evaluating its effects particularly valuable. Affirmative action improves the outcomes of targeted students. Specifically, race-based quotas raise the share of black students in federal universities, an effect not observed with income-based quotas alone. Affirmative action has downstream positive consequences for labor market outcomes. The results suggest that income and race-based quotas beneficiaries experience substantial long-term welfare benefits. There is no evidence of mismatching or negative consequences for targeted students' peers.  
\end{abstract}



\section{Introduction}
\label{sec:Introduction}

This survey aims to improve the understanding of affirmative action policies in higher education by analyzing the design and implementation of placement quotas for underrepresented minority (URM) students in Brazil. In that country, public universities are tuition-free, there is a national centralized university admission system, and, for federal universities, both income and racial-based quotas have been mandatory since 2012. The staggered reforms in the higher education system in Brazil, which include a centralized admissions system for most public universities and transparent implementation of affirmative action, improve our understanding of the costs and benefits of affirmative action in higher education. In particular, we can investigate how alternative designs, such as income-based quotas, can improve rules based on racial criteria. 

According to earlier reviews on the subject [eg. \cite{holzer2000assessing}, \cite{fryer2005affirmative}, \cite{long2007affirmative}, and  \cite{arcidiacono2016affirmative}], some of the central gaps in the empirical evidence regarding affirmative action consist of estimating individual outcomes and answering questions regarding the optimal design of such policies. As to outcomes, the main issues are related to the extensive margin (whether individuals attend college at all) or the intensive margin (e.g., effort and job market outcomes). On the optimal design of affirmative action policies, are there viable alternatives to race-based quotas, such as income-based rules or "percent plans," which provide automatic admission to public universities for students above a given percentile in class or state rank? 

The Brazilian experience is informative because scattered initiatives coalesced into a standardized policy for all federal universities with staggered implementation. The resulting system is transparent, with public knowledge about students admitted through affirmative action. Candidates must declare if they are competing for general or reserved seats. Rich datasets allow for robust identification strategies and statistical power. In particular, we summarize the evidence on the effects of affirmative action policies on the following dimensions: who gets in, the performance of targeted students, effort, and labor market outcomes. In addition, we demonstrate that income-based quotas are not a substitute for race-based affirmative action and detail how verification committees, used for limiting moral hazard, work. Methodologically, the present article uses extant empirical research on affirmative action in Brazil with one caveat: articles must have clear identification strategies, regardless of their use of observational or other types of data.  

For instance, researchers have exploited the use of both income and race as admission criteria in Brazil to conclude that race-based affirmative action has increased the enrollment of black students. In contrast, race-neutral policies have failed to do so \citep{vieira2019affirmative}. Such findings are consistent with results in the U.S.; \cite{ellison2021efficiency} document that race-neutral policies in Chicago are less efficient than plans that explicitly consider race. 

The effect of affirmative action is more pronounced in the most competitive and prestigious courses, which see the most increase in the share of students enrolling through affirmative action policies. In addition, there is no evidence of a mismatch between targeted students and majors; overall academic achievements don't change, and the gap between the GPAs of targeted students and other students is reduced as graduation approaches.

Moreover, there are improvements in labor market outcomes for the beneficiaries, although gains are not uniform. There is no indication of reduced welfare for students other than the ones displaced by the quotas.  There is also evidence that many quota students who complete their studies would not have had an alternative path to higher education. 

Evidence for the country supports a view that nationally coordinated policies cannot be disregarded as ineffective in meetings the goals of affirmative action (the ethos of the decentralized American system follows from an implicit distrust of national policies). In Brazil, income, race, and disability-based affirmative action coexist for most public institutions, while private colleges can choose to adopt similar systems or eschew them altogether. In particular, the evidence that income-based quotas alone are insufficient for increasing student body diversity is novel. Another result is the documentation that the reduction in transaction costs may be a significant low-hanging fruit for policymakers that aim to increase minorities' applications to higher education.

The remainder of this paper is organized as follows. Section 2 briefly introduces the origins of affirmative action policies in Brazil. Section 3 describes the institutions and empirical evidence involving affirmative action in Brazil. Section 4 concludes. 

\section{Antecedents and operational details of affirmative action in tertiary education in Brazil}
\label{sec:affActionInBrazil}

Educational affirmative action can be divided broadly between race-based, ethnicity-based, and income-based policies. Race-based affirmative action in tertiary education emerged in the 1960s in the United States. It is decentralized across regions and institutions (For instance, California banned affirmative action in 1996, and voters in 2020 rejected overturning the ban.)

The goals for race-based affirmative action are clear: to improve access to tertiary education by underrepresented minorities (URMs), which would eventually lead to improved labor, health, and financial outcomes. \cite{arcidiacono2016affirmative} argue that it is challenging to identify the returns to college majors and major plus college combinations. Nevertheless, the dual goals of affirmative action of generating a diverse student body and supporting the educational attainment of minority students are complementary in the absence of strong mismatching effects. Mismatch is the potential situation in which a policy makes students worse off by attending a more selective university relative to attending a less selective school. 

In the remainder of this section, we present a quick background of higher education in Brazil before affirmative action and describe the two major educational reforms in the country's higher education system. Mechanically, affirmative action increases the enrollment of minority candidates, but we are most interested in the details of displacing and displaced students.

\subsection{Antecedents of affirmative action in Brazil}
\label{subsec:affirmativeaction}

In 1950, only 5\% of white Brazilians and 0.5\% black and mixed Brazilians had graduated from high school \citep{andrews2014racial}. Until the 1990s, things did not change much, with only a small elite having access to higher education. In 1992, only 7.4\% (white) and 1.5\% (black and mixed) of all young adults were enrolled in a tertiary education institution. Today, there are approximately 4 million students in the higher education system, which comprises only 17\% of all young adults.   

We can see inequality in access to higher education on several dimensions. By 2000, only 6.8\% of adults older than 25 held a college degree (and only 0.4\% a postgraduate degree), but for black and mixed Brazilian, that share was only 2.2\%. According to the latest Census (in 2010), the share of college enrollment for individuals aged 18 to 22 in the poorest families (lower income quartile) was 3.7\%, and for families in the top quartile, 34.2\%. Even though access to higher education has improved in the last few decades, the repressed supply of Brazilians with a higher education degree has kept returns to education high in the country. \cite{souza2019estimating} find that the average return to education---measured as the increase in hourly wages per year of education---in Brazil has been at least 11.1\% since 2000 (around 14\% from 1995 to 2003, 11.1\% from then until 2014.)

In the past, access to higher education, private and public, was decentralized. Candidates undertook university-specific exams. Time constraints, registration fees, and travel costs limited the number of institutions one could aim for. For instance, one of the authors vividly remembers not undertaking the exam for one of Rio's three prominent public (and, most importantly at the time, free!) universities due to the exam fees and transportation costs.

The Brazilian government enacted two significant reforms in the early 2010s: the centralization of the placement system and the dissemination of affirmative action in the country.

Now, individuals choose the institution and major while comparing their grades with cutoff scores, while in the previous decentralized system, choices about which institutions a student would apply to had to be made months before the institution-specific admission exam \citep{mello2023affirmative}. It is unclear, in principle, who would benefit more from this increased access.

In Brazil, many forces have created a conducive environment for the state to tackle racial discrimination: the universalization of primary education in the early 1990s, macroeconomic stabilization, and structural changes in the supply of tertiary education. Hyperinflation was vanquished in 1994, and many private universities started expanding in the late 1990s. In some respects, public controversies in Brazil about the merits, fairness, and practical logistics of affirmative action policies are similar to the American debate. However, the design of affirmative action policies is unique to the country. 

Before the two major reforms in the early 2000s, admissions have been solely based on the candidates' results in university-specific standardized tests. Therefore, affirmative action policies consist of additional layers on top of this criterion. However, some policy choices, such as the verification commissions, have no counterpart in the U.S. \citep{lempp2019eyes}.

Brazil's first affirmative action policy for university entrance exams emerged in 2000 when the Rio de Janeiro state government introduced legislation (Bill 3,524/2000), which required state universities to reserve 50\% of the seats for public school students. In 2001, through Bill 3,708/2001, local legislators established racial quotas, with 40\% of the public school seats reserved for black and mixed-race Brazilians. In 2004, Universidade Nacional de Brasília became the first federal university to follow suit. 

Affirmative action spread through the country amidst a broader education reform program that replaced university-specific exams with a centralized clearinghouse that allocates students to federal higher education institutions using a national exam. Student matching is done through a dynamic centralized procedure \citep{bo2022iterative,machado2021centralized}. Below we present the operational details of affirmative action in the country.

\subsection{Operational details of affirmative action in higher education in Brazil}
\label{subsec:detailsAffAcBrazil}

The affirmative action law established that the seats in each program offered by federal universities for each incoming major class should be split into four groups in four layers.

Students must declare if they are competing for reserved or general seats. Seats are equally divided into general and reserved seats. For instance, if the university opens 100 seats for a certain major,  students declare if they are competing for the 50 general competition spots or one of the 50 reserved seats.   

\begin{figure}[t]
    \centering
    \includegraphics[scale=0.75]{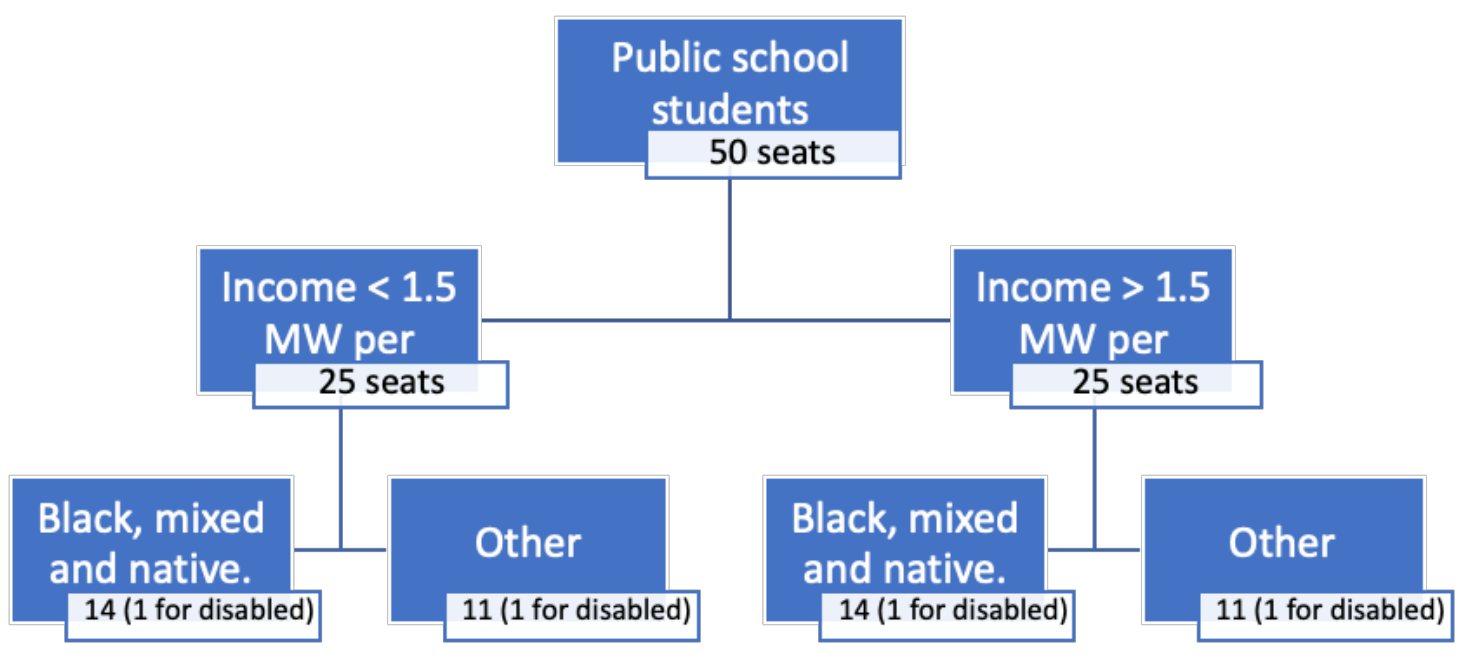}
    \caption{Distribution of quotas}
    \label{fig:quotas}
\end{figure}

Of the four layers for the reserved seats, represented in figure \ref{fig:quotas}, the first layer divides students by their high school of origin; 50\% of the seats are reserved for those from public high schools. 

In the second layer, students are classified by family income. Lower-income students are defined as those from households with an average monthly income below 1.5 times the national minimum wage; moreover, 50\% of the quota seats are reserved for lower-income students.

In the third layer, the subdivision is minority status (students who self-identify as black, mixed, or indigenous Brazilians). The fourth layer is for students with disabilities. Students may choose not to declare their position in each of these layers. If candidates don't announce they are competing for reserved seats, they are placed in the general category.

Rules are uniform for all federal universities in Brazil. State universities, funded by local governments, are not obliged to follow the same system but usually do. For instance, candidates for the State University of Sao Paulo (USP) have two options: they can sit through an entrance exam for the institution (FUVEST) or use their grades from the national exam (ENEM). In 2022, of the 8,000 places, 3,000 have been reserved for candidates who want to use their ENEM grades. In Rio de Janeiro, two of the three state universities select students solely through the national centralized system, while UERJ has a two-tier system similar to USP. Most (if not all) state institutions have affirmative action policies similar to that of federal universities.

\subsection{The unstable consensus for affirmative action in Brazil}
\label{subsec:consensus}

The process was led by many activists and social organizations that challenged the conventional wisdom on Brazil's perceived "racial democracy." In 1995, Fernando Henrique Cardoso's government created the Inter-ministerial Working Group for the Black Population, comprised of members of government ministries and social movements \citep{bailey2018support}.

But it was under Luiz Inácio Lula da Silva's government, which created the Special Office for the Promotion of Racial Equality, that affirmative action policies started to be enshrined into law \citep{heringer2015affirmative}. From 2004, when the Education Ministry proposed authorizing federal universities to adopt quotas, to 2012, when the law required that federal universities adopt affirmative action, a surprising majority of Brazilians strongly supported these policies \citep{bailey2018support}, even though many groups opposed them. 

Regarding the legislative debate, through the 1990s and 2000s, a consensus strong enough for new legislation to gain steam emerged slowly. \cite{frade2020translation} describes how the process included a transformation of context (common sense achieved by society) into legal discourse (law), employing the categorization of everyday concepts into legal concepts. In 2004, the Brazilian Supreme Court unanimously decided that affirmative action policies are legal. The court considered its transitory nature implicit in any legislation with reserved places for underrepresented minorities. Initially, the 2012 legislation establishing affirmative action in higher education in Brazil was supposed to be revised in 2022. Bill number 12,711 has established income and race-based quotas, and Bill 13,409, from 2016, created quotas for students with disabilities. Also, in 2016, legislation changed the revision process; unless Congress passes new rules clarifying the competence for revising affirmative action in Brazil, the 2012 bill will remain in effect.

\section{Empirical evidence from the dissemination of affirmative action in Brazil}
\label{sec:Empiricalevidence}

The present survey aims to include the most extant research on the topic. We search via multiple keywords in Portuguese and English to acquire as many studies as possible. We discard studies without empirical content or clear identification strategies, regardless of their use of observational or other data types. 

Before proceeding with the evaluation of affirmative action policies, we describe racial disparities in Brazil. Then, we answer one of the central questions in this study: how do targeted students perform vis-a-vis general competition students? Finally, we summarize the impact of affirmative action on the market and non-market outcomes for targeted students. Market evidence, as usual, is restricted to labor and educational outcomes. In addition, we highlight the possible long-term repercussion of affirmative action by tracing shifts in norms and attitudes, including issues such as do quotas change the effort by nonwhite Brazilians in preparation for entrance exams and during their higher education studies.

\subsection{Racial disparities}
\label{subsec:RacialDisparities}

Brazil is the world's last country to formally abolish slavery, a historical blemish that still impacts nonwhites in the country today. Some racial dynamics in the country mirror the U.S., but one crucial difference makes the analysis of affirmative action in Brazil even more appealing. A long literature on the significance of skin color in Brazil discusses how racial disparities arise despite perceived inclusion in the horizontal dimension \citep{telles2014race} and no documentation of overt racial conflict \citep{roth2016race}. \cite{telles2014race} argues that high rates of intermarriage and low levels of residential segregation indicate inclusion that is distinct from in the U.S. (a difference along a "horizontal" dimension) but that does not impact indicators of material wealth, social status, and educational attainment (the "vertical" dimension). That matters because it implies that desegregation and other means of "inclusion" are unlikely to be enough to reduce indicators in the vertical dimension.

We present the results from a few studies to situate the reader regarding Brazil's differences in the vertical dimension between white and nonwhites. Racial disparities might arise from non-market interactions, tastes, attitudes, and statistical discrimination. Direct and indirect effects on health outcomes and racial attitudes are potentially self-reinforcing \citep{reskin2012race}. We start with general indirect outcomes and move to wage and education attainment differences. 

\cite{pavao2013autoavaliaccao} weigh the effects of racial discrimination on three health outcomes: depression, physical disorders, and well-being self-evaluation. They interview 3,863 black and mixed households and measure racial discrimination on a self-reported 9-level scale. Any previous experience of racial discrimination results in a 37\% higher risk of worse perceived health outcomes; in addition to overall effects, physical disorders are also more likely (55\% higher), as is depression (77\%). Complementing this, \cite{faerstein2014race} have analyzed data for over 3.000 civil servants (of whom 48\% are black). They also rely on the self-reported experience of racially charged events. Of the African-Brazilians in their sample, 16\% have reported facing episodes of racial discrimination. For them, high blood pressure is 40\% more likely than other African-Brazilians. Finally, racial discrimination is associated with increased odds of common mental disorders, according to \cite{bastos2014age}.

Regarding racial attitudes, \cite{lima2004sucesso} run an experiment in which white Brazilians are asked to evaluate pictures of both financially successful and poor individuals. Successful black Brazilians are perceived as whiter, and observers ascribe more human qualities to the whiter individuals in the photographs. Machado de Assis (1839-1908), one of Brazil's most celebrated writers, "got whiter" as he became more famous \citep{flynn2013whiter}; newspapers doctored his pictures to tone down his skin color. Something similar has happened with the depiction of President Barack Obama in the U.S. \citep{kemmelmeier2014biases}. 

Perception leads to reality. In Brazil, the nonwhite wage gap has been estimated by \cite{cornwell2017wage}. They take advantage of the fact that almost all employers in Brazil are required to report race, alongside other employee characteristics, on the annual employment-population survey. The authors identify the cases in which workers change jobs and are classified differently by the new employer, from nonwhite to white or vice-versa. As much as 40 percent of the racial wage gap remains after controlling for all individual characteristics. Workers classified as white on the destination job (from nonwhite in the origin job) see an average wage gain of 1.7 percent. On average, workers who make the reversed racial journey are expected to lose 2.1 percent of their income. For those reported as white in the origin and destination jobs, the white wage premium is approximately 7 percent. Results are corroborated by \cite{gerard2021assortative}, who find an unexplained preference for white workers at higher-paying establishments, and \cite{reis2017fields}, who brings evidence that disparities in the distribution of racial groups across fields of study help explain 18\% of the total median earnings differential in 2000 and 33\% in 2010 (yet another study with similar results is \citealp{hirata2017evidence}). \cite{gerard2021assortative} also show that nonwhites are less likely to be hired by high-wage firms, which explains about 20 percent of the racial wage gap for both genders, and that firm-specific pay premiums for nonwhites are compressed, contributing another 5 percent for that gap. 

Racial inequalities are also present in education markets. \cite{portella2019health} observe black-white education gaps of half a year of schooling by the time children are 11 years old, growing to 1.3 years by the time individuals turn 22. They also find that gaps are formed early in life and that family income, maternal education, number of prenatal visits, maternal smoking during pregnancy, and maternal mental health explain it. \cite{fernandes2017brazilian} complements these results by highlighting the role of school segregation; nonwhite students are under-represented in private schools. To disentangle the role played solely by race, \cite{marteleto2016racial} use a within-family twin approach to assess whether racial disparities in education exist between twins of different races and whether such inequality varies by gender. They find that "even under this stringent test of racial inequality, the nonwhite educational disadvantage persists and is especially pronounced for nonwhite adolescent boys" (p.1185). This disadvantage is again compounded because nonwhite adolescents have a penalty of –0.13 years of schooling compared with their white peers. Racial differences early in life appear to drive racial disparities later in college academic performance \citep{francis2012busing} Finally, what interests us the most is the most apparent evidence regarding higher education outcomes from Francis and \cite{francis2012aredistributive}. First-difference regressions on pairs of siblings reveal that black identity and gender hurt university entrance exam scores \citep{francis2016light}.

Many other studies show that racial disparities are commonplace in Brazil; levels of racial inequality are higher in Brazil than in the United States \citep{andrews2014racial}. Still, since our goal is not to present the origins of racial disparities in the country (see \cite{almeida2019racismo} for a throughout discussion on the subject), we concentrate on how affirmative action has changed the student body composition in higher education in Brazil.

\subsection{Who gets in? The evolution of access to higher education in Brazil}
\label{subsec:WhoGetsIn}

As we have seen, the two primary higher education reforms in Brazil have been the centralized admissions system (SISU, which is relatively transparent \citep{hoxby2015high}) and the law requiring standardizing affirmative action to all federal universities. Thus, it is essential to disentangle the role played by each reform in shaping the number and performance of URMs in Brazilian public universities.

Regarding centralization, \cite{machado2021centralized} exploit time variation in the institutions' adoption of  the clearinghouse to investigate student sorting, migration, and enrollment impacts. They find higher geographical mobility and turnover of seats by admitted students, which indicate that students' search is intensified. Overall, the system is more efficient than the previous decentralized admissions process. For instance, students who would find it costly to apply to several universities may now consider any institution in the country. Students, including those who may benefit from affirmative action, have access to information on how competitive admission to all universities in the country is, improving their choices regarding majors and institutions. 

\cite{mello2023affirmative} answers that and other questions by building on the cross-sectional and time variation of the progressive adoption of both policies and dividing the enrollment effect into three kinds: two for each policy separately and one for its interaction. The benefits from the centralized system accrue disproportionately to students from wealthier families, crowding out low-income groups from the least competitive degrees. However, affirmative action increases low-income individuals' enrollments mechanically and through behavioral responses. The author finds a complementary effect that protects low-income groups from crowding out of centralization.\footnote{``Results show that the full adoption of AA—from 0 to 50 percent of reserved vacancies—increases enrollments of  public school (PS),  public school  non-White (PSNW) and '  public school  low-income students (PSLI) by, respectively, 9.9, 7.0, and 2.4 percentage points, an increase of 18, 29, and 34 percent for the average program. The full adoption of SISU acts in the opposite direction, decreasing enrollments of these groups by 3.8, 2.8, and 4.1 percentage points, representing a negative effect of 7, 12, and 59 percent for the average program. Finally, the interaction between both policies creates an additional effect that increases the enrollments of all vulnerable groups.'' \citep[p.~168]{mello2023affirmative}.} Interestingly, mobility constraints only seem to affect low-income students in majors with low expected future returns. 

Centralization and affirmative action significantly enhance low-income and minority students access. Previously, it would be almost impossible for poor students to compete for seats in universities from other states. More importantly, the most significant enrollment changes have been those with the lowest share of minority students; courses that are competitive, selective, prestigious, and, therefore, with the greatest potential for job market returns \citep{senkevics2019perfil,estevan2019bcan}. The 2012 quota policy for federal universities in Brazil not only had a mechanical effect on the inclusion of the target audience but also generated incentives for students from the beneficiary groups to compete for places, possibly revealing a repressed demand for higher education \citep{senkevics2019perfil}. Complementing the study by \cite{mello2023affirmative} and using data for a very selective university, \cite{estevan2019bcan} find an increased likelihood of minority students applying for and getting accepted to more prestigious majors. A further indication of the repressed demand hypothesis is the effect of the education reforms on disabled candidates. \cite{dalcin2020unintended} find that centralizing the application process diminishes mobility costs more salient for individuals with a disability. They find that the percentage of entrants with a disability in institutions that have adopted the Sisu after seven years was 0.63 percentage points higher than in those outside the system, almost doubling the share of entrants with disabilities (only 0.77\% before the reform).

So far, the picture is that affirmative action in Brazil has more than mechanically enhanced student body diversity, and there is no initial evidence of changes in dropout rates. Still, we must now answer the question: how do targeted students perform, and are there negative spillovers to other students?

\subsection{Performance of targeted students}
\label{subsec:PerformanceTargeted}

One of the primary concerns of opponents of affirmative action is the mismatching hypothesis, which posits that many of the beneficiaries of preferences are so misplaced academically that they would be better off in the absence of affirmative action. \cite{arcidiacono2016affirmative} show that there are clear positive average effects of college quality on a host of outcomes in the U.S., but that mismatch effects could potentially still arise. The authors also argue that the net impact of racial preferences comes down to whether the strength of the overall college quality effect is larger or smaller than any matching effects.

A necessary condition for mismatch to be of concern is that targeted students must perform poorly. Unless targeted students are misplaced academically, they couldn't be better off without affirmative action policies that increase the share of URMs in higher education institutions. In Brazil, a common concern at the height of the debate regarding affirmative action policies is that affirmative action might reduce the overall quality of education because targeted students would either slow down classes or fail to keep up with the contents of lectures.

There is clear evidence that the beneficiaries of affirmative action perform well, to some extent even better than expected, URMs keep up with the material and do not slow down classes. For instance,  \cite{francis2012busing} employ a robust identification strategy based on a difference-in-difference estimation to test for the performance of targeted students. They have access to comprehensive administrative data (albeit for a single institution), which they complement by interviewing 799 students. Students' performance is based on GPAs. Affirmative action does not widen racial disparities for the most selective courses and departments. More importantly, the authors can compare GPAs for displacing and counterfactually displaced students in these selective courses (as GPAs for actually displaced students are obviously not available). Throughout their studies, targeted students close any initial grade gap relative to other students. There is no change in the performance of other students. By the time they graduate, the GPAs of targeted students are not significantly different from that of other students. There is one caveat. Before the quotas are introduced, there are modest racial disparities in GPA among students in selective departments. Still, the introduction of racial quotas does not seem to impact these in any way. 

\cite{valente2017performance} complement these results and find that students admitted to public universities under the federal affirmative action law perform similarly to students who enter through the regular competition exam. Their study uses observations from national exams of over 1 million students to regress performance indicators on some individual characteristics. In addition, the authors document that quota students in private universities perform slightly better than students admitted through traditional methods. They also do not find evidence of mismatching. Most importantly, there is no indication of negative externalities on other students' performance. Other observational studies with results are \cite{wainer2017politicas} and \cite{de2019avaliaccao}. The former finds that students who receive a hardship stipend score higher than their classmates (\citealp{dynarski2018closing} find a similar effect in the U.S.). Using propensity score matching, the latter does not find meaningful differences between the performance of targeted and non-targeted students.

When it comes to dropout rates, the evidence also points towards rejecting the mismatching hypothesis. Using data from all federal universities which implemented quota policies during intakes from 2010 to 2013, \cite{honoratoEtAL2022} show that, when considering first and second-year dropout rates, targeted students presented a dropout rate of 9\%, compared to 10\% of non-targeted students. \cite{pena_matos_coutrim_2020} find no statistically significant difference when comparing the dropout rate of targeted and non-targeted students admitted at the Federal University of Ouro Preto. Finally, \cite{costa2020beyond} describe pre and post-reform dropout rates for Brazil's second-largest federal university and show that neither overall nor race-specific dropout rates changed much over time.

\cite{oliveira2022affirmative} analyze the margins of adjustment for beneficiaries of affirmative action in a top university in Brazil, where prospective students must choose a major before the entrance test and cannot switch it while in college. Dropout rates are not higher than in other settings, such as the United States. Targeted students fail more courses early, but there is evidence of strong catching up for students who do not drop out. They reduce the number of credit hours taken in the first and second college years but compensate by taking more credit hours in the final years. Drop-out rates are small (5\%) in their setting of the initial years of affirmative action in the country. \footnote{One could consider the possibility that these results are affected by changes in grading methods on the part of the teachers in response to an influx of students who are not as well prepared for university. The lack of impact on non-targeted students' dropout rates combined with the catching up observed by \cite{oliveira2022affirmative}, however, would require the grade inflation to target specifically quota students and phase out during their studies. No known systematic change in grading policy would have resulted in this. On the other hand, if there was grade inflation in the entire cohort, we would expect an increase in graduation rates of non-quota students, which was not found in these studies.}

 Other studies that corroborate these results with quasi-natural experiments are conducted for single institutions. Regardless of the design, the studies broadly agree that targeted students learn as much as other students, do not reduce overall educational quality, and conclude their studies similarly to non-targeted students. As there is no evidence that beneficiaries of affirmative action are misplaced academically, there is no direct mechanism for mismatch to be of much concern in the Brazilian context. 
\cite{ribeiro2016affirmative} is an example of a single-institution study. She investigates the performance of targeted students in the most prestigious law school in the State of Rio de Janeiro using students' bar examination performance. She presents significant results comparing targeted and non-targeted students. She finds that students who enter the university through affirmative action have an 11.5 and 7.7 percentage points higher chance of passing the two phases of the bar exam than the pool of other students with similar entrance scores. However, when the sample of students is restricted to those who pass the first exam phase, URMs' exam scores are four percentage points lower. To some extent, underperformance (other examples include \cite{araujo2020diferencial} and \cite{cavalcanti2019desempenho}) is expected, as entrance cutoffs for targeted students are lower. However, there is clear evidence that targeted students benefit substantially from access to this elite major. Public school quota beneficiaries whose university entrance exam scores are close to the admittance cutoff see their approval rate rise by 52 percentage points. In addition, displaced candidates (tracked through their enrollment in other institutions) do not experience any drop in their bar exam approval rate, nor do the colleagues of the targeted students. \cite{motte2020effect} find similar results after they use a regression discontinuity for students of a federal university. They also document another driver of affirmative action on students' performance. In Brazil, it is common for first-year classes to be staggered. Students with the highest grades start in the first semester. At the Federal University of Bahia, the authors use the discontinuity between students narrowly allocated into the first and second semesters. The beneficiaries of affirmative action that come last among the best first-class students obtain a positive impact on their labor market earnings, probably due to the networking created during their extra time at the university. Nevertheless, the opposite effect is found for non-quota students; performance is higher if they start in the second semester.

In a country where returns to education are significant, one would expect sizable benefits from attending college. Any potential mismatch between skills and selected majors could diminish those effects. However, outcomes from Brazil suggest that mismatch may not be much of a concern. There seems to be no mechanism that the standardized affirmative action policies for federal universities in Brazil hurt its purported beneficiaries.

So far, we have seen that affirmative action in Brazil seems to work as intended, for the most part. Still, we must answer a few questions: does affirmative action change effort and labor market outcomes? Are there alternative designs? Can race-blind income-based quotas replace race-based affirmative action?

\subsection{Effort and labor market outcomes}
\label{subsec:EffortMarketOutcomes}

We have seen that enrolled students do not seem to be affected by affirmative action. One question remains, however. What about pre-university and college efforts? Recent models have posited that it is unlikely that affirmative action reduces effort. For instance, in the \cite{cotton2020affirmative} model, the impact of preferential policies differs across population groups; it matters whether one is a beneficiary of the policy and one's relative ability within the group. The model shows that a preferential treatment policy that targets disadvantaged students may increase their effort by placing them within reach of outcomes that would otherwise be unattainable. Still, affirmative action could generate negative externalities if other high-school students significantly reduce their effort. Yet there is no evidence for that, either for URM or other students, in the Brazilian context.

Three critical studies estimate effort in Brazil, one for a state university and another for a federal university, both using quasi-natural experiments and an observational one for the entire country. 

Regarding pre-university efforts, \cite{estevan2019aredistribution} investigate a system that awards bonus points to prospective quota applicants in the admission exam of a state university in São Paulo. They find a sizable redistribution of university access: almost 10\% of all admitted applicants would not have been accepted without the policy. Yet, there is no difference in the preparatory effort of targeted and non-targeted students. Moreover, there is no evidence of significant heterogeneity in behavior for students above or below the admission cutoff grade or their relative distance from this cutoff. They suggest that behavioral responses are too small to be detectable in the data if they occur at all. They posit that students across demographic groups realize the value of higher education and generally have equal levels of aspiration (for evidence regarding the U.S. context, see \citealp{oyserman2017seeing}.)

\cite{francis2012busing} confirm these results by estimating effort with data from over 2,000 students and 24,000 applicants on two admissions cycles from 2003-2005 for another institution that introduced race-based quotas. They find that students successfully admitted through the quota system are from families with significantly lower socioeconomic status. In addition, racial quotas did not reduce the pre-university effort of either applicants or students. This study is complemented by \cite{pelegrini2022there}, who find indirect evidence of a higher effort by targeted students in a sample of 130,728 students. This indirect evidence is based on documentation that differences between quota and non-quota students diminish as students approach graduation. That confirms the results in the previous section that targeted students perform well during their undergraduate studies. In addition, there is no indication that non-targeted students change their behavior. These bodies of evidence, while limited, suggest that white and wealthier students, following the introduction of affirmative action policies, do not change their behavior towards preparation for the entrance exams meaningfully. 

Next, we consider the issue of labor market outcomes of targeted students. \cite{francis2018black} use a regression discontinuity design to connect the admissions outcomes of high-performing applicants of a single institution in 2004–2005 to their education and labor market outcomes in 2012. They find that enrolling at that particular university through affirmative action raises the likelihood that quota applicants work as a director or manager and in a public sector job. Quota applicants end up with significantly more years of education than non-attending applicants. More importantly, they are more likely to graduate from college.

Although effort does not seem to change, there may be high costs to the students displaced by affirmative action. Displaced students are, as expected, white and from higher socioeconomic status \citep{francis2012aredistributive}. The authors find that 39.7\% of displacing and 19.0\% of displaced applicants had family income equal to or less than R\$ 1,500 (60\% above the minimum wage at the time), whereas 8.5\% of displacing and 30.7\% of displaced applicants had family income greater than R\$ 5,000. Unfortunately, there is only one source on labor market outcomes regarding displaced students, and even then, the evidence is just for one law school in Brazil. In this study, displaced students do not experience any drop in their bar exam approval rate \citep{ribeiro2016affirmative}.

These studies indicate no significant behavioral change for targeted and non-targeted candidates after the major affirmative action reform. That matters because candidates who believe their chances have diminished might adjust their behavior accordingly.

\subsection{Racial vs. income-based quotas}
\label{subsec:RacialVsIncomeQuotas}

Racial discrimination in Brazil is widespread. Still, that does not mean racial quotas are the most efficient way to improve minority access to tertiary education. Income-based quotas may serve the purpose of alleviating racial disparities while easing concerns about fairness to the poorest white Brazilians. There is one prevailing myth about affirmative action in Brazil. Most poor Brazilians are nonwhite, so income-based quotas would improve diversity without free-riding from wealthier nonwhite individuals.

Brazilian policymakers have responded to the tension between income and race-based affirmative action by establishing a four-layer quota system, with race and income as the main criteria. However, \cite{vieira2019affirmative} find that universities that adopt race-blind policies see no significant changes in the racial profile of their admitted students. In contrast, those that use race-based quotas see more substantial increases in the participation of nonwhites than alternative arrangements. From 2004-2012, universities could design affirmative action policies, and many introduced race-blind policies, while others introduced both race-blind and racial quotas. \cite{vieira2019affirmative} design difference-in-difference estimates of race-blind and racial preference policies using the staggered adoption of different approaches. They use a large sample of first-year students from all Brazilian federal universities during the period. Since they have control groups, estimates account for unobserved shocks concurrent with adopting affirmative action. Before policy adoption, the trends in the shares of students from disadvantaged groups are parallel across treated and controlled university programs.

More importantly, the largest effect of increasing the proportion of nonwhites from racial quotas on the student body's racial composition is in highly competitive programs. In addition, the authors do not find unintended consequences of race-based quotas affecting the gender composition of academic programs. This finding is compatible with \cite{emerson2017colorblind}, which analyzes colorblind policies in the U.S., and shows that they do not seem to increase the diversity of student bodies. 

While racial quotas successfully increase cohort diversity, race-blind methods don't. The reason may be that aggregate observations such as "most minorities are low-income" do not translate into URMs benefiting from low-income quotas.  When it comes to who gets in and who gets out, details about correlations and students' preferences, which are not in these aggregate statistics, might lead to an "unintuitive" consequence of these policies \cite{aygun2021college}.

\subsection{Verification committees}
\label{subsec:VerificationCommittees}

Black identity is not a binary variable, but the quota system treats it as such. The bureaucratic constraints on implementing affirmative action require substantiation for the students who self-declare for reserved seats. In the U.S., admissions officers are responsible for this process, but in Brazil, given that the system has been centralized, the government has established verification committees for implementing affirmative action.

The functioning of committees is regulated by an ordinance published on April 6, 2018. Committees are composed of five members and five alternates, and their compositions are publicly available. Training is required, and members should be of different genders, ethnicity, and geographical origin. 

Candidates indicate if they want to compete for reserved places or not. To minimize work, the committees determine eligibility for three times the number of approved candidates, which means that members only meet after exams are over and grades are determined. Candidates must meet the committee, which means that the process is not based solely on secondary information. Interviews must be filmed, but the footage is only used if candidates appeal the initial determination by the committee. Eligibility is supposed to follow one, and only one, criterion based on phenotype characteristics. Committees do not, and are not supposed to, determine a candidate's ethnicity and must be careful about not making public judgments about candidates' racial identity; the committees' goal is merely to establish eligibility for that particular exam. Eligibility is valid for that specific exam. There is evidence that committees reject approximately 10\% of applicants' claims to be considered for reserved seats \citep{filho2019}. 

In Brazil, there is a sense that individuals share a "social gaze" that allows them to determine who is black and who is not. Verification committees have evolved to deal with the tension between the idea of a generalized social gaze—due to which "everybody" knows who is black in Brazil—and the emphasis on a skilled gaze one has to learn and train to decide who is entitled to a quota place \citep{lempp2019eyes}.

There are multiple possible sources of errors and fraud. However, the nature of a centralized system limits the possibility of direct or indirect bribery, such as with the recent scandals in the U.S. where parents paid university officials to get their kids into selective colleges. In July 2020, Universidade de Brasilia expelled 15 students for violating the affirmative action process \citep{otoboni2020}. Some types of behavior infringe upon the spirit of the system, while others are standard instances of fraud. Some families may shift income or provide false reports for their children to compete for income-based reserve seats. Others may enroll their children in two high schools, one public and one private, so that their children can have the option to declare for either the reserved or general competition seats.  

However, the frequency of fraud is mitigated by the fact that candidates have their basic information and the provisional determination of their eligibility made public, for better or worse. Civil society, including social media warriors and NGOs, has publicized cases of individuals claiming to be white but who have been approved for a reserved spot for minorities in other contexts.

Unfortunately, to the best of our knowledge, few articles have analyzed, and even then only tangentially, the effectiveness of the verification committees. \cite{francis2012busing} and \cite{francis2013endogenous} find that the introduction of racial quotas has induced some students to misrepresent their racial identity but inspired other individuals, those who are the darkest-skinned, to consider themselves black. Nevertheless, it is a reasonable assumption that misrepresentation, never a large share of total quota applications in the first place, should decrease over time as members of the committees learn, and the applicants' expectations are moderated.

\section{Conclusion}
\label{sec:conclusion}

This article argues that the widespread use of affirmative action policies in Brazil allows for studying its effects to a greater extent than in most settings. 

We find evidence indicating that the Brazilian experience can be informative for the future of affirmative action in Brazil and other countries. Before exploring the empirical literature that documents the outcomes of preferential access policies in Brazil, we contextualize the Brazilian environment by describing racial disparities in the country, focusing on health, education, and labor market outcomes. Racial and income disparities provide the impetus for affirmative action in the country. We describe the implementation mechanism of affirmative action in federal universities in Brazil, with its four-layer centralized system that differentiates students by income, race, and disability status. 

The surveyed empirical evidence indicates that targeted students are not worse off under affirmative action. In addition, there is no indication of a change in the effort by potentially displaced students or, for that matter, by targeted students. There is also no indication of mismatching. 

Overall, affirmative action works, in the Brazilian context, as intended: it increases diversity, generates net benefits for targeted students, and does not reduce student welfare unless quotas directly displace them. In addition, targeted students do not under-perform and there is no indication of negative externalities on the performance of other students. 

The Brazilian evidence supports the view that nationally coordinated policies can efficiently implement affirmative action policies, even in the face of multiple layers of student selection. Other countries can choose to replicate parts of the Brazilian system without worrying about widespread downsides. More importantly, the Brazilian experience shows that income and race-based quotas are not substitutes and that candidates not targeted by these policies don't reduce their effort, at least in the context of a country where returns to education are substantial. However, there are some caveats. Some studies lack the most up-to-date identification strategies, and there is little information on displaced students' outcomes. The existing evidence does not point toward severe effects, but we cannot rule out the possibility that displaced students incur high costs.

On the issue of possible distortions, there are also some misrepresentations of racial identity, and the verification commissions create transaction costs that wouldn't otherwise exist.

\bibliographystyle{ecca}
\bibliography{quotasSurvey}

\end{document}